\documentclass[11pt]{article}

\usepackage{latexsym} \usepackage{epsf}
\usepackage{a4}
\usepackage{amsfonts}

\def\be{\begin{equation}}
\def\ee{\end{equation}}
\def\bea{\begin{eqnarray}}
\def\eea{\end{eqnarray}}
\def\nn{\nonumber}
\def\dd{\partial}
\def\'{\prime}
\def\a{\alpha}
\def\b{\beta}

\def\o{\omega}

\def\e{\varepsilon}
\def\ba{\begin{array}}
\def\ea{\end{array}}

\begin{document}
\vspace{3cm}
\begin{title}
\bf{\LARGE{\bf{ Polynomial Realization of $s\ell_q(2)$ and Fusion
Rules at Exceptional Values of $q$}}}
\end{title}
\vspace{3cm}

{\begin{center}

{\large
D.~Karakhanyan$^a$\footnote{e-mail: karakhan@lx2.yerphi.am} \&
Sh.~Khachatryan$^a$\footnote{e-mail: shah@moon.yerphi.am} \\ [3mm] }
\end{center}
}

\begin{itemize}
\item[$^a$]
Yerevan Physics Institute , \\
Br.Alikhanian st.2, 375036, Yerevan , Armenia.
\end{itemize}
\vspace{3cm}

\begin{abstract}
Representations of the $s\ell_q(2)$ algebra are constructed in the
space of polynomials of real (complex) variable for $q^N=1$. The
spin addition rule based on eigenvalues of Casimir operator is
illustrated on few simplest cases and conjecture for general case
is formulated.
\end{abstract}
\vspace{3cm}

\vspace{1cm}

{\bf{keywords}}: {\it{Quantum groups, Fusion rules}.}

\newpage
\section{Introduction}

The quantum groups were invented \cite{f} by L.D.Faddeev and the
Leningrad school on inverse scattering method in order to solve
integrable models. Quantum groups have links with mathematical
fields such as Lie groups, algebras and their representations,
special functions, knot theory operator algebras, non-commutative
geometry and many others and have a lot of interrelations with
physics: quantum inverse scattering method, theory of integrable
systems, conformal and quantum field theory, etc \cite{Ar1,dck,
GRAS,GRS,Jim}. It is expected that quantum groups will lead to a
deeper understanding of the concept of symmetry in physics. The
quantum symmetry bears close similarity with non-deformed
classical symmetries, especially in field of representation
theory. However that similarity occurs only at general values of
deformation parameter and ends for so called exceptional values of
$q$ ($q^N=1$).

Degeneracy in the case when $q$ is given by a root of unity is
accompanied by enlarging of the center of symmetry group and
changing the structure of representation space of theory.
Physically it expressed in the fact that XX model has more wide
symmetry than XXZ Heisenberg model.

The new features in representation theory which appear under
quantum deformation of Lie groups with parameter $q$ ($q^N=1$)
were studied by V.~Pasquier and H.~Saleur \cite{PS} who introduced
the notion of indecomposable representation, by D.~Arnaudon
\cite{Ar1} who classified all representations of quantum groups by
types $\mathcal{A}$ and $\mathcal{B}$ and some other authors
\cite{dck}.

In this article we propose an explicit realization of
representations in the space of polynomials, which is very useful
in practical calculations. This approach is especially fruitful in
the context of Universal R-matrix \cite{kk}. Such approach allows
to give an explicit operator realization of the Universal R-matrix
in the space of polynomials. Based on this approach it is possible
to give a heuristic illustration of basic regulations appearing in
case $q^N=1$.

\section{$s\ell_q(2)$ algebra and co-product}

The representation of algebra generators as linear function of
derivatives acting on polynomials of a certain number of variables
is known in mathematical literature \cite{k}. Physicists are more
familiar, however, with the matrix representation and it seems
more useful to explain the relation between them.

The simplest differential realization of the Lie algebras is built
on the homogeneous polynomials. The case of $s\ell(2)$ algebra is
the most convenient for pedagogical purposes. The fundamental
representation is given by Pauli matrices:\be\label{fund}
S^+=\left(\ba{cc}0&1\\0&0\ea\right),\quad
S^-=\left(\ba{cc}0&0\\1&0\ea\right),\quad S^z=\frac
12\left(\ba{cc}1&0\\0&-1\ea\right),\ee corresponding doublet
representation is \be\label{r2}
R_2=\{\;\;\mid\uparrow\rangle=\left(\ba{cc}1\\0\ea\right),
\;\mid\downarrow\rangle=\left(\ba{cc}0\\1\ea\right)\},\ee The
Hermitean conjugation has to be introduced in order to formulate a
non-trivial orthogonality condition for these elements. We
consistently work only with the ket-states avoiding introduction
of the Hermitean conjugation and bra-states. Since we would like
to construct a realization on polynomials, we introduce two
independent variables, $x$ and $y$ and identify them with elements
of the doublet representation, \be
\mid\uparrow\rangle=y,\quad\mid\downarrow\rangle=x.\ee Then
(\ref{fund}) and (\ref{r2}) imply
$$
S^+x=y,\qquad S^+y=0
$$
\be S^+y=0,\qquad S^-y=x,\ee
$$
S^+x=-\frac 12x,\qquad S^+y=\frac 12y.
$$
These relations immediately allow to realize the generators in the
differential form: \be S^+=y\dd_x,\quad S^-=x\dd_y,\quad S^z=\frac
12(y\dd_y-x\dd_x).\ee Now one can construct all higher
representations of the type: \be\label{rj}
R_j=(\underbrace{R^{(2)}\times R^{(2)}\times\ldots\times
R^{(2)}}_j)_{symmetrized}.\ee This is very important general
feature of all differential realizations with the group generators
which are linear in derivatives: if realization is found for some
representation, it can be applied to the product. Indeed, for any
representation $R_j=R_1\times R_2\times\ldots$ the action of
generators $S^a$, by definition, must reduce to:
$$
S^aR_j=(S^aR_1)\times
R_2\times\ldots+R_1\times(S^aR_2)\times\ldots+\ldots,
$$
which exactly coincides with the standard rules of
differentiation. This property holds due to linearity of
generators with respect to derivatives. The symmetrization
mentioned above is obtained automatically because all doublets are
identical. In $s\ell(2)$ case symmetrized product (\ref{rj})
covers all possible representations. This is not so for higher
rank groups.

The elements of (\ref{rj}) are the homogeneous expressions of form
$x^ky^{n-k}$, where $n$ is number of factors in (\ref{rj}).

Homogeneous realization of generators is not very suitable. More
convenient is inhomogeneous one: \be (R^{(2)}\times
R^{(2)}\times\ldots\times R^{(2)})=\{x^n,\,x^{n-1}y,\,x^{n-2}y^2,
\,\ldots , y^n\}\equiv x^n\{1,\,\xi,\,\xi^2,\,\dots
,\,\xi^n\},\qquad \xi=\frac yx.\ee Now we can rewrite $s\ell(2)$
generators so that they will act on $\{1,\,\xi,\,\xi^2,\,\ldots
,\,\xi^n\}$ thus mixing different powers of $\xi$:
$$
\!\!S^+x^n\{(1,\xi)\times(1,\xi)\times\ldots\}\!=\!(S^+x^n)
\{(1,\xi)\times(1,\xi)\times\ldots\}+x^n\!\sum_{k=1}^{n}
\{(1,\xi)\times \ldots\times(T^+(1,\xi))
\times\ldots\times(1,\xi)\}
$$
\be =nx^n\xi\{(1,\xi)\times(1,\xi)\times\ldots\}+
x^n\!\sum_{k=1}^{n} \{(1,\xi)\times \ldots\times((0,-\xi^2))
\times\ldots\times(1,\xi)\},\ee and similar relations for $S^-$
and $S^z$.

Omitting the overall factor $x^n$ one can deduce from these
relations that \be S^+=n\xi-\xi^2\frac\dd{\dd\xi},\quad S^z=-\frac
n2+\xi\frac\dd{\dd\xi},\quad S^-=\frac\dd{\dd\xi}.\ee Here $j=n/2$
has sense of spin of the representation.

This realization can be extended to the case of other symmetry
groups as well as to the case of deformed symmetry $s\ell_q(2)$.
The $s\ell_q(2)$ algebra is defined by generators $e=S^-$,
$f=S^+$, $k=q^{2S^z}$, $k^{-1}=q^{-2S^z}$ and commutation
relations:
$$
kfk^{-1}=q^2f, \quad kek^{-1}=q^{-2}e,\quad
kk^{-1}=1=k^{-1}k,\quad [f,e]=\frac{k-k^{-1}}{q-q^{-1}}.
$$
The scaling or dilatation symmetry, generated by $S^z$ survives
under this deformation. Hence polynomial structure of the lowest
weight representations can be kept as well as explicit form of
$S^z$. Another two generators then are given by following
expressions:

\textit{Definition 1.$\;$ $\{ Homogeneous \}$} \be S^z=\frac
12(y\dd_y-x\dd_x),\quad S^-= x/y\frac{q^{\frac
12(y\dd_y-x\dd_x)}-q^{\frac 12(x\dd_x-y\dd_y)}}{q-q^{-1}}, \quad
S^+= y/x\frac{q^{\frac 12(x\dd_x-y\dd_y)}-q^{\frac
12(y\dd_y-x\dd_x)}}{q-q^{-1}}\,,\ee and the representation space
is given by the set of homogeneous expressions
$\{x^{n-k}y^k\}_{k=0}^n$. It is also possible to pass to
inhomogeneous representation $\{\xi^k\}_{k=0}^n$. The generator of
translations $S^-$ goes to the q-derivative. Remaining generators
are also given by finite-difference operators and we have come to

\textit{Definition 2. $\;$ $\{ Inhomogeneous \}$} \be
e=S^-=\frac{\xi^{-1}}{q-q^{-1}}(q^{\xi\dd_\xi}-q^{-\xi\dd_\xi}).\ee

$$
k=q^{2\xi\dd_\xi-n},\qquad
f=S^+=\frac{\xi}{q-q^{-1}}(q^{n-\xi\dd_\xi}-q^{\xi\dd_\xi-n}).
$$
Here we follow the notations and conventions ref.\cite{Ar1} and
define the co-product to be:
$$
\Delta(k)=k\otimes k,
$$
\be \Delta(e)=e\otimes 1+k\otimes e,\ee
$$
\Delta(f)=f\otimes k^{-1}+1\otimes f,
$$
which is a little bit simpler than usual definition
$$
\Delta(q^{S^z})=q^{S^z}\otimes q^{S^z},
$$
\be \Delta(S^\pm)=S^\pm\otimes q^{\pm S^z}+q^{\mp S^z}\otimes
S^\pm,\ee except for the transformation rules under exchange
$q\leftrightarrow kq^{-1}$. This rule allows an unambiguous
definition of $s\ell_q(2)$ generators for higher tensor products
as well. Let one has triple tensor product. Then there exist two
possible definitions:
$$
k=k_{12}k_3,\quad e=e_{12}+k_{12}e_3,\quad f=f_{12}k^{-1}_3+f_3,
$$
and
$$
k=k_1k_{23},\quad e=e_1+k_1e_{23},\quad f=f_1k^{-1}_{23}+f_{23},
$$
however result in both cases is the same:
$$
k=k_1k_2k_3,\quad e=e_1+k_1e_2+k_1k_2e_3,\quad
f=f_1k^{-1}_2k^{-1}_3+f_2k^{-1}_3+f_3.
$$
The indices here denote the numbers of the spaces. In tensor
product notations they would be rewrite as

$$
 {\rm {if} }\;\; \Delta(A)=\sum_{i} B_{i}\otimes
C_{i},\qquad A, B_{i},C_{i} \in s\ell_q(2),\;\; {\rm{ then} }\quad
\Delta\Delta(A)\!=\!\sum_{i}\! \Delta (B_{i})\otimes C_{i}\!=\!
\sum_{i} B_{i}\otimes \Delta(C_{i}).
$$
This relation reflects on the co-associativity property of the
co-product for Hopf algebras. \be (\Delta \otimes 1)\Delta=(1
\otimes \Delta)\Delta. \ee Then it is obvious by induction that
result is consistent for higher tensor products too.

\section{Representations of $s\ell_q(2)$ at exceptional values
of deformation parameter.}

The representations of $s\ell_q(2)$ at general values of $q$ have
the same structure as in non-deformed (classical) case. However
when $q$ takes exceptional values, i.e. is given by roots of unity
($q^N=1$), there appear many differences. The center of
$s\ell_q(2)$ is enlarged for these values of $q$: in addition to
conventional quadratic Casimir operator \be \label{casim}
\mathcal{C}=fe+(q-q^{-1})^{-2}(q^{-1}k+qk^{-1})=
ef+(q-q^{-1})^{-2}(qk+q^{-1}k^{-1}), \ee there appear also new
Casimirs:
$$
e^{\mathcal{N}},\qquad f^{\mathcal{N}},\qquad
k^{\pm{\mathcal{N}}},\qquad {\mathcal{N}}=\left\{\ba{cc}
{\mathcal{N}}=N, \quad N\;\;
 {\rm{is\;\; odd}}\\
{\mathcal{N}}=\frac 12 N,\quad N\;\; {\rm {is\;\;
even}}\ea\right.,\qquad q^{\mathcal{N}}=\pm 1.
$$
It is not hard to establish the following relation:
\be\label{constr}
f^{\mathcal{N}}e^{\mathcal{N}}=\prod_{n=0}^{\mathcal{N}-1}\left(
\mathcal{C}-\frac{q^n-2+q^{-n}}{(q-q^{-1})^2}\right)+\frac{(k^{
\mathcal{N}}-1)(k^{-\mathcal{N}}-1)}{(q-q^{-1})^{2\mathcal{N}}}.\ee
There appear \cite{Ar1} the representations of new $\mathcal{B}$
type or cyclic ones, which have no classic counterpart. They
defined as \be e^{\mathcal{N}}\neq 0,\qquad \mbox{or/and}\quad
f^{\mathcal{N}}\neq 0.\ee Here we are interested in $\mathcal{A}$
type or lowest weight representations, which also differ from
corresponding ones for general values of $q$. They defined as \be
e^{\mathcal{N}}=0,\qquad f^{\mathcal{N}}=0,\qquad
k^{\mathcal{N}}=\pm 1.\ee In general case representation is
parameterized by eigenvalues of four Casimir operator, which are
related by constraint (\ref{constr}) and one can formulate the
following

\textit{Proposition.} The most general differential expression for
generators consistent with $s\ell_q(2)$ algebra contains three
arbitrary parameters: \be\label{abl} f\equiv
S^+=q^{\lambda/2}x\frac{q^{\a-x\dd}-q^{x\dd-\a}} {q-q^{-1}},\quad
e\equiv S^-=\frac{q^{-\lambda/2}}x
\frac{q^{x\dd-\b}-q^{\b-x\dd}}{q-q^{-1}}, \ee
$$
k\equiv q^{2S^z}=q^{-\a-\b}q^{2x\dd}.
$$
This differential realization is very convenient to describe
lowest weight representations: there always exists lowest weight
vector $\o_0=1$. It annihilated by lowering generator $e$. Other
vectors of representation can be obtained by repeatedly acting of
rising generator $f$. Now, it becomes obvious that dimension of
such representation then bounded by $\mathcal{N}$, because after
$\mathcal{N}$ steps it will repeat itself due to
$f^{\mathcal{N}}\sim 1$. So it consists of not more than
$\mathcal{N}$ elements, otherwise it will be reducible. Generally
irrep. of type $\mathcal{A}$ with spin $j$ has dimension $2j+1$.
Indeed substituting (\ref{abl}) into lowest and highest weight
conditions $e\cdot 1=0$ and $f\cdot x^{2j}=0$ one can obtain that
general form of generators of spin $j$ representation are:

\textit{Definition 3.}
\be\label{genj}e_{(j)}^\mathcal{A}=\e_{(j)}x^{-1}\frac{q^{x\dd}-
q^{-x\dd}}{q-q^{-1}},\quad f_{(j)}^\mathcal{A}=x\frac{q^{2j-x\dd}-
q^{x\dd-2j}}{q-q^{-1}},\quad
k_{(j)}^\mathcal{A}=\e_{(j)}q^{2x\dd-2j},\ee and representation
space is given by polynomials which powers do not exceed $2j$. We
shall denote it $P_{j}$.

\section{Tensor product of spins one and one half representations
and fusion rules.}

The co-multiplication of quantum algebra enables us to define
tensor product of representations.
\\

$\bullet$ Consider at first tensor product of two representations
of spin $\frac 12$ in polynomial spaces with variables $x_1,x_2$.
According to the definition of co-product $s\ell_q(2)$ generators
in this case take the form:
$$
e=\frac\e{q-q^{-1}}\left[\frac
1{x_1}(q^{x_1\dd_1}-q^{-x_1\dd_1})+\frac{q^{2x_1\dd_1-1}}{x_2}
(q^{x_2\dd_2}-q^{-x_2\dd_2})\right],
$$
\be\label{gen2} f=\frac 1{q-q^{-1}}\left[
q^{1-2x_2\dd_2}x_1(q^{1-x_1\dd_1}-q^{x_1\dd_1-1})+x_2
(q^{1-x_2\dd_2}-q^{x_2\dd_2-1})\right],\ee
$$
k=\e q^{2x_1\dd_1+2x_2\dd_2-2},
$$
and act in $P_{\frac 12\frac 12}=\{1,\ x_1,\ x_2,\ x_1x_2\}$ as
follows:
$$
e\cdot 1=0,\quad e\cdot x_1=\e,\quad e\cdot x_2=\e q^{-1},\quad
e\cdot x_1x_2=\e(qx_1+x_2)
$$
\be k\cdot 1=\e q^{-2}\cdot 1,\quad k\cdot x_i=\e x_i,\quad
i=1,2,\quad k\cdot x_1x_2=\e q^2x_1x_2, \ee
$$
f\cdot 1=(qx_1+x_2),\quad f\cdot x_1=x_1x_2,\quad f\cdot x_2=
q^{-1}x_1x_2,\quad f\cdot x_1x_2=0
$$
Then one deduces that Casimir operators act on these vectors
according to the formulae:
$$
e^2=0=f^2, \quad k^2=1
$$
\be \mathbb{C}\cdot 1=\e\frac{q^3+q^{-3}}{(q-q^{-1})^2},\quad
\mathbb{C}\cdot\e x_1=(qx_1+x_2)+\e\frac{q^3+q^{-3}}
{(q-q^{-1})2}x_1, \ee
$$
\mathbb{C}\cdot x_2=\e(x_1-qx_2)+\e\frac{q+q^{-1}}
{(q-q^{-1})^2}x_2,\quad \mathbb{C}\cdot
x_1x_2=\e\frac{q^3+q^{-3}}{(q-q^{-1})^2}x_1x_2.
$$
So one can see that tensor product of two spin $\frac 12$
representations decomposes according to eigenvalues of Casimir
operator $\mathbb{C}$:
$$
c_1=\e\frac{q^3+q^{-3}}{(q-q^{-1})^2},
$$
on triplet of vectors: $\{1,\;\; (x_1+q^{-1}x_2),\;\; x_1x_2\}$
and
$$
c_3=\e\frac{q+q^{-1}}{(q-q^{-1})^2},
$$
on singlet $x_1-qx_2$. That means that tensor product of two spin
$\frac 12$ decomposes in direct sum of spin one and spin zero for
any $N$ ($q^N=1$) except for $N=4$, which corresponds to XX
Heisenberg model. So for $N\neq 4$ spin addition rule is not
deformed: couple of spins $\frac 12$ decomposes to spin one and
spin zero. Indeed spin one representation can be obtained as
symmetrized part of mentioned above tensor product by setting spin
zero component $x_1-qx_2=0$, i.e. $x_2=q^{-1}x_1$, then
representation space consists of quadratic polynomials of one
variable $x=x_1$: $P_{\frac 12\frac 12}^{sym}\sim P_1$. Then
derivative $\dd_2$ vanishes and generators take the form:
$$
e=\frac\e{q-q^{-1}}(q^{x\dd}-q^{-x\dd}),\qquad f=\frac
1{q-q^{-1}}(q^{2-x\dd}-q^{x\dd-2}),\qquad k=\e q^{2x\dd-2},
$$
standard for representation of spin one.

The case $N=4$ require separate consideration. When deformation
parameter takes values $q=\pm i$ degeneracy of Casimir's
eigenvalues takes place, ($c_1=c_2=0$) and vectors of triplet and
singlet are unified into one multiplet. Moreover, two eigenvectors
which are linear with respect to $x$'s ($x_1+qx_2$ and $x_1-q^
{-1}$) coincide each to other when $q=\pm i$. In other words the
eigenvectors of Casimir operator do not longer form a basis in
$P_{\frac 12\frac 12}$ and have to be completed by one additional
vector. The visual evidence of representations unification in this
case is provided by matrix representation:
setting \\
$\o_1=1=\!\left(\!\ba{cccc}0\\0\\0\\1\ea\!\right)\!,\;\;$
$\o_2=\frac
12(qx_1+x_2)=\!\left(\!\ba{cccc}0\\0\\1\\0\ea\!\right)\!,\;\;$
$\o_3=x_1+qx_2=\!\left(\!\ba{cccc}0\\1\\0\\0\ea\!\right)\!,\;\;$
$\o_4=x_1x_2=\!\left(\!\ba{cccc}1\\0\\0\\0\ea\!\right)\!,$ one
obtains that in this basis operator $\mathbb{C}$ has form:
\be\label{c4}
\mathbb{C}=\e\left(\ba{cccc}0&0&0&0\\0&0&0&0\\0&1&0&0\\0&0&0&0\ea\right),
\ee which obviously cannot turn to diagonal form by linear
transformation of basis. In other words representation space
$I^{(4)}$ cannot be decomposed on invariant subspaces according to
eigenvalues of Casimir operator $\mathbb{C}$. Representation
$I^{(4)}$ is called indecomposable, because it is neither
reducible (generators mix representation vectors and do not leave
any invariant subspace) nor irreducible (being initially
introduced as tensor product).
\\

$\bullet$ Consider next tensor product of three representations.
It has form:
$$
e=\frac\e{q-q^{-1}}\left[\frac 1{x_1}(q^{x_1\dd_1}-
q^{-x_1\dd_1})+\frac {q^{2x_1\dd_1-2j_1}}{x_2}(q^{x_2\dd_2}-
q^{-x_2\dd_2})+\right.
$$
$$
\left. +\frac {q^{2x_1\dd_1+2x_2\dd_2-2j_1-2j_2}}
{x_3}(q^{x_3\dd_3}-q^{-x_3\dd_3})\right],
$$
\be\label{3rep} k=\e q^{2x_1\dd_1+2x_2\dd_2+2x_3\dd_3-3},\ee
$$
f=\frac 1{q-q^{-1}}\left[q^{2j_2+2j_3-2x_2\dd_2-2x_3\dd_3}x_1
(q^{2j_1-x_1\dd_1} -q^{x_1\dd_1-2j_1}) +\right.
$$
$$
\left. +q^{2j_3-2x_3\dd_3}x_2(q^{2j_2-x_2\dd_2}-q^ {x_2
\dd_2-2j_2})+ x_3(q^{2j_3-x_3\dd_3}-q^{x_3\dd_3-2j_3})\right].
$$
First consider the case $j_1=j_2=j_3=\frac 12$. On representation
vectors of $P_{\frac 12\frac 12\frac 12}$ Casimir operator acts as
follows:
$$
\mathbb{C}\cdot 1=\e\frac{q^4+q^{-4}}{(q-q^{-1})^2}\cdot 1,\quad
\mathbb{C}\cdot x_1=\e (q^2x_1+qx_2+x_3+\frac{q^2+q^{-2}}
{(q-q^{-1})^2}x_1),
$$
$$
\mathbb{C}\cdot x_2=\e (qx_1+x_2+q^{-1}x_3+\frac{q^2+q^{-2}}
{(q-q^{-1})^2}x_2),\quad \mathbb{C}\cdot x_3=\e
(x_1+q^{-1}x_2+q^{-2}x_3+\frac{q^2+q^{-2}} {(q-q^{-1})^2}x_3),
$$
\be
\label{eig3}
\mathbb{C}\cdot x_1x_2=\e ((1+q+\frac{2}
{(q-q^{-1})^2})x_1x_2+qx_1x_3+x_2x_3),
\ee
$$
\mathbb{C}\cdot x_1x_3=\e
(qx_1x_2+2(1+\frac{1}{(q-q^{-1})^2})x_1x_3+q^{-2}x_2x_3),
$$
$$
\mathbb{C}\cdot x_2x_3=\e
(x_1x_2+q^{-1}x_1x_3+(1+q^{-2}+\frac{2}{(q-q^{-1})^2})x_2x_3),\quad
\mathbb{C}\cdot x_1x_2x_3=\e\frac{q^4+q^{-4}}{(q-q^{-1})^2}\cdot
x_1x_2x_3,
$$
>From these relations one can deduce that Casimir has two
eigenvalues \be {{c}}_1=\e\frac{q^4+q^{-4}}{(q-q^{-1})^2}, \ee on
quartet of vectors:
$$
\o_0\equiv 1,\quad \o_1\equiv (x_1+q^{-1}x_2+q^{-2}x_3),\quad
\o_2\equiv (x_1x_2+q^{-1}x_1x_3+q^{-2}x_2x_3),\quad \o_3\equiv
x_1x_2x_3
$$
and \be {{c}}_2=\e\frac{q^2+q^{-2}}{(q-q^{-1})^2}, \ee on another
quartet:
$$
\varphi_1^{(1)}\equiv(x_1-qx_2),\quad
\varphi_1^{(2)}\equiv(x_1-q^2x_3),\quad
\varphi_2^{(1)}\equiv(x_1x_2-qx_1x_3),\quad
\varphi_2^{(2)}\equiv(x_1x_2-q^2x_2x_3).
$$
Now it can be easily checked that: \be e\o_0=0,\qquad e\o_1=\e(1+
q^{-2}+q^{-4})\o_0,\qquad e\o_2=\e(q+q^{-1})\o_1,\qquad
e\o_3=q^2\o_2, \ee
$$
f\o_0=q^2\o_1,\qquad f\o_1=(q+q^{-1})\o_2,\qquad f\o_2=(1+
q^{-2}+q^{-4})\o_3,\qquad f\o_3=0.
$$
and \be e\varphi_1^{(1)}=0,\quad e\varphi_1^{(2)}=0,\quad
e\varphi_2^{(1)}=\e q^{-1}(\varphi_1^{(2)}-\varphi_1^{(1)}), \quad
e\varphi_2^{(2)}=\e(q-q^{-1})\varphi_1^{(1)}+\e
q^{-1}\varphi_1^{(2)},\ee
$$
f\varphi_1^{(1)}=q^{-1}(\varphi_2^{(2)}-\varphi_2^{(1)}),\quad
f\varphi_1^{(2)}=(q-q^{-1})\varphi_2^{(1)}+q^{-1}\varphi_2^{(2)},
\quad f\varphi_2^{(1)}=0,\quad f\varphi_2^{(2)}=0.
$$
>From these relations one can deduce that first quartet corresponds
to four dimensional representation of spin $\frac 32$, while
second one constitutes two spin $\frac 12$ representations, when
$\mathcal{N}>3$. In other words, when $q$ is a root of unity
higher degree tensor product of three spin one half representation
decomposes in the same way as for general values of $q$ or in a
classical case.

The low values of $\mathcal{N}$ require separate
consideration.

When $\mathcal{N}=2$, one has $e^2=0=f^2$ on vectors $\o_i$ and
$\varphi^{(a)}_b$ due to $q+q^{-1}=0$ and mentioned vectors can be
combined into pairs: $(\o_0,\o_1)$, $(\o_2,\o_3)$,
$(\varphi_1^{(1)},\phi_2)$, $(\varphi_2^{(1)} ,\phi_1)$, which
constitute four spin $\frac 12$ representation spaces. Here we
denoted $\phi_1=\varphi_1^{(2)}-\varphi_1^{(1)}$ and
$\phi_2=\varphi_2^{(2)}-\varphi_2^{(1)}$. So the tensor product of
three spin $\frac 12$ decomposes for $q=\pm i$ into the sum of
four spin $\frac 12$ irreps.

Now let $\mathcal{N}=3$, i.e. $N=3$ or $N=6$. Then one has $k^2=1$
and $e^3=0=f^3$ due to $1+q^{-2}+q^{-4}=0$ and eigenvalues
of Casimir operator ${\mathbb{C}}$ become degenerate:
$$
c_1=c_2=-\frac{\e \cos(\frac{4\pi k}3)}{2\sin^2(\frac{2\pi k}3)}
=\e\frac{q^2+1+q^{-2}-1}{q^2+1+q^{-2}-3}=\frac\e 3,
$$
due to $q^2+1+q^{-2}=0$ and one can easily check that triplets
$\{\o_1, \varphi_1^{(1)}, \varphi_1^{(2)}\}$ and $\{\o_2,
\varphi_2^{(1)}, \varphi_2^{(2)}\}$ become linearly dependent and
two quartets of vectors are unified into one multiplet - sextet, which
is an indecomposable representation like a quartet in the case $q=\pm
i$. Eigenvectors (\ref{eig3}) do not longer form a basis in representation
space and have to be completed by another two vectors.
It is easy to establish relations: \be \mathbb{C}(x_1+\a
x_2+\b x_3)=\frac\e 3(x_1+\a x_2+\b
x_3)+(q^2+q\a+\b)(x_1+q^{-1}x_2+q^{-2}x_3),\ee and
$$
\mathbb{C}(x_1x_2+a x_1x_3+b x_2x_3)=
$$
$$
=\frac\e 3(x_1x_2+a x_1x_3+b
x_2x_3)+(q^2+qa+b)(x_1x_2+q^{-1}x_1x_3+q^{-2}x_2x_3).
$$
Now one can see that upon "symmetrization" with respect
$1\leftrightarrow 2$ or $1\leftrightarrow 3$ "antisymmetric"
doublet $(\o_1^{(2)}), \o_2^{(2)}$: ($x_1-qx_2$, $x_1x_3-qx_2x_3$)
or ($x_1-q^2x_3$, $x_1x_2-q^2x_2x_3$), corresponding to spin
$\frac 12$ representation decouples:
$$
e\o_1^{(2)}=0,\quad f\o_1^{(2)}=\o_2^{(2)},\quad
e\o_2^{(2)}=\e\o_1^{(2)},\quad f\o_2^{(2)}=0,
$$
while remaining six vectors are unified into an indecomposable
representation $I_1^{(6)}=\{\o_i^{(6)}\}$: \be \o_1^{(6)}=1,\quad
\o_2^{(6)}=x_1+q^{-1}x_2 +q^{-2}x_3,\quad \o_3^{(6)}=x_1+qx_2
+q^2x_3,\ee
$$
\o_4^{(6)}=x_1x_2+q^{-1}x_1x_3 +q^{-2}x_2x_3,\quad
\o_5^{(6)}=x_1x_2+qx_1x_3 +q^2x_2x_3,\quad \o_6^{(6)}=x_1x_2x_3.
$$
Representing these vectors as columns $ (\o_i^{(6)})^j=\delta_i^j$
one obtains that Casimir operator $\mathbb{C}$ acts as $6\times 6$
matrix: \be\mathbb{C}=\frac\e
3+3q^2\left(\ba{cccccc}0&0&0&0&0&0\\0&0&0&0&0&0\\0&1&0&0&0&0\\
0&0&0&0&0&0\\0&0&0&1&0&0\\0&0&0&0&0&0\\\ea\right),\ee which again
cannot be turned to diagonal form. Hence corresponding sextet
$I_1^{(6)}$ is an indecomposable representation. $I_1^{(6)}$ can
be obtained directly as tensor product of spins $\frac 12$ and 1
while in present case one has one additional doublet: $(\{\frac
12\}\otimes\{\frac 12\})\otimes\{\frac
12\}=(\{1\}\oplus\{0\})\otimes\{\frac
12\}=\{I_1^{(6)}\}\oplus\{\frac 12\}$. So one can deduce that
representations with spins higher than 1 (i.e. 3/2 etc.) are not
allowed when $\mathcal{N}=3$. According to classification given by
D.~Arnaudon in \cite{Ar1} there exists another indecomposable
sextet, which appears in tensor product of two spin 1
representations or in the quartic product of spins $\frac 12$.
\\

$\bullet$ For the product of two spin 1 representations one can
obtain:
$$
e=\frac\e{q-q^{-1}}\left(\frac
1{x_1}(q^{x_1\dd_1}-q^{-x_1\dd_1})+\frac{q^{2x_1\dd_1-2}}
{x_2}(q^{x_2\dd_2}-q^{-x_2\dd_2})\right),\quad k=\e
q^{2x_1\dd_1+2x_2\dd_2-4},
$$
\be f=\frac 1{q-q^{-1}}\left(x_1q^{2-2x_2\dd_2}
(q^{2-x_1\dd_1}-q^{x_1\dd_1-2})+
x_2(q^{2-x_2\dd_2}-q^{x_2\dd_2-2})\right),\ee On representation
space $P_{11}$ Casimir operator $\mathbb{C}$ acts as follows:
$$
\mathbb{C}\cdot 1=\e\frac{q^5+q^{-5}}{(q-q^{-1})^2},\quad
\mathbb{C}x_1=\e\left((q+q^{-1})(q^2x_1+x_2)+\frac{q^3+q^{-3}}
{(q-q^{-1})^2}x_1\right),
$$
$$
\mathbb{C}x_2=\e\left((q+q^{-1})(x_1+q^{-2}x_2)+\frac{q^3+q^{-3}}
{(q-q^{-1})^2}x_2\right),
$$
$$
\mathbb{C}x_1^2=\e(q+q^{-1})\left(q^2x_1^2+(q+q^{-1})x_1x_2+
\frac{x_1^2} {(q-q^{-1})^2}\right),
$$
\be \mathbb{C}x_1x_2=\e\left(q^2x_1^2+x_2^2+2(q+q^{-1})x_1x_2+
\frac{q+q^{-1}}{(q-q^{-1})^2}\right),\ee
$$
\mathbb{C}x_2^2=\e(q+q^{-1})\left(q^{-2}x_2^2+(q+q^{-1})q^{-2}x_1x_2+
\frac{x_2^2} {(q-q^{-1})^2}\right),
$$
$$
\mathbb{C}x_1^2x_2=\e(q+q^{-1})\left(x_1x_2^2+(1+q^2)x_1^2x_2+
\frac{x_1^2x_2}{(q-q^{-1})^2}\right),
$$
$$
\!\!\!\mathbb{C}x_1x_2^2=\e(q+q^{-1})\!\left(x_1^2x_2+(1+q^{-2})x_1x_2^2+
\frac{x_1x_2^2} {(q-q^{-1})^2}\right)\!,\quad \mathbb{C}\cdot
x_1^2x_2^2=\e\frac{q^5+q^{-5}}{(q-q^{-1})^2}x_1^2x_2^2.
$$
Using these relations one can deduce that $\mathbb{C}$ has in
general only three eigenvalues:
$$
c_1=\e\frac{q^5+q^{-5}}{(q-q^{-1})^2},
$$
on vectors
$$
\{\varphi^{(1)}_i\}_{i=1}^5=\{1,\ (x_1+q^{-2}x_2),\ (
x_1^2+q^{-3}( q+q^{-1})^2x_1x_2+ q^{-4}x_2^2),\ (
x_1^2+q^{-2}x_1x_2^2),\ x_1^2x_2^2\},
$$
$$
c_2=\e\frac{q^3+q^{-3}}{(q-q^{-1})^2},
$$
on the set
$$
\{\varphi^{(2)}_\a\}_{\a=1}^3=\{(x_1-q^2x_2),\
(x_1^2+(q^{-3}-q)x_1x_2-x_2^2),\ x_1^2x_2-q^2x_1x_2^2\},
$$
and third eigenvalue
$$
c_3=\e\frac{q+q^{-1}}{(q-q^{-1})^2},
$$
on combination
$$
\{\varphi^{(3)}\}=x_1^2-(q+q^{-1})x_1x_2+q^2x_2^2.
$$
According to these relations tensor product of two spin one
representations decomposes on quintet, triplet and singlet:
$\{1\}\otimes\{1\}=\{2\}\oplus\{1\}\oplus\{0\}$ as it takes place
for general values of deformation parameter. As we already learned
above, the case when $q$ is root of unity of low degree has to be
studied carefully, because eigenvalues $c_i$ of Casimir operator
$\mathbb{C}$ become degenerate. The simplest case $\mathcal{N}=2$
i.e. $q=\pm i$ is now excluded because spin 1 representation is
not allowed for these $q$.

When $\mathcal{N}=3$ i.e. $q^3=1$ or $q^6=1$ one sees that
$e^3=0=f^3$, $c_1=c_3=\frac\e 3$ and quadratic with respect to
$x_i$ eigenvectors become linearly dependent:
$$
x_1^2+(q^{-3}-q)x_1x_2-x_2^2=x_1^2-(q+q^{-1})x_1x_2+q^2x_2^2,
$$
due to $q^{-2}+q^{-4}=-1$ and $q^{-4}=q^2$ when $q^6=1$. The
triplet $\{\varphi^{(2)}_\a\}_{\a=1}^3$ corresponding to spin 1
decouples. Adding to the set $\{\varphi^{(1)}_i\}_{i=1}^5$ one
more quadratic with respect to $x_1,$ $x_2$ vector
$\varphi^{(1)}_6=x_1^2+(2-\frac 32 q)x_1x_2+(\frac 12
q^{-2}-3q^{-3})x_2^2$ to complete basis in representation space
one obtains another indecomposable sextet
$I^{(6)}_2=\{\varphi^{(2)}_i\}_{i=1}^6$. The Casimir operator
after exchange $\varphi^{(1)}_6\leftrightarrow\varphi^{(1)}_4$
acts on this set as $6\times 6$ matrix: \be\mathbb{C}=\frac\e
3-(\frac 56+q)\left(\ba{ccccccc}0&0&0&0&0&0\\0&0&0&0&0&0\\0&0&0&0&0&0\\
0&0&1&0&0&0\\0&0&0&0&0&0\\0&0&0&0&0&0\\\ea\right),\ee i.e. it
again cannot be made diagonal by linear transformation of vectors
of representation.

Another value of deformation parameter which leads to degeneracy
of Casimir's eigenvalues corresponds to $q^8=1$. In this case one
obtains relations: $e^4=0=f^4$, $k^2=1$, $c_1=c_2=-\frac {\sqrt
2}2=-c_3$. Then corresponding vectors become linearly dependent
and has to be completed to form basis of an eight-dimensional
indecomposable representation $I^{(8)}_1$.
\\

$\bullet$ Let us now turn to the tensor product of four
representations of spin $\frac 12$. Generators, acting on
$P_{\frac 12 \frac 12 \frac 12 \frac 12}$ can be represented in
following form:
$$
f=\frac 1{q-q^{-1}}\left[q^{3-2x_4\dd_4-2x_3\dd_3-2x_2\dd_2}
x_1(q^{1-x_1\dd_1}-q^{x_1\dd_1-1})+q^{2-2x_4\dd_4-2x_3\dd_3}x_2
(q^{1-x_2\dd_2}-q^{x_2\dd_2-1})+\right.
$$
$$
\left.+q^{1-2x_4\dd_4}x_3 (q^{1-x_3\dd_3}-q^{x_3\dd_3-1})
+x_4(q^{1-x_4\dd_4}-q^{x_4\dd_4-1})\right],
$$
\be k=\e q^{2x_1\dd_1+2x_2\dd_2+ 2x_3\dd_3+2x_4\dd_4-4},\ee
$$
e=\frac\e{q-q^{-1}}\left[\frac
1{x_1}(q^{x_1\dd_1}-q^{-x_1\dd_1})+\frac{q^{2x_1\dd_1-1}}{x_2}
(q^{x_2\dd_2}-q^{-x_2\dd_2})+\right.
$$
$$
\left.+\frac{q^{2x_1\dd_1+2x_2\dd_2-2}}{x_3}
(q^{x_3\dd_3}-q^{-x_3\dd_3})+\frac{q^{2x_1\dd_1+2x_2\dd_2+
2x_3\dd_3-3}}{x_4}(q^{x_4\dd_4}-q^{-x_4\dd_4})\right].
$$

Casimir operator acts on $P_{\frac 12 \frac 12 \frac 12 \frac 12}$
as follows:
$$
\mathbb{C}\cdot 1=\e\frac{q^5-q^{-5}}{(q-q^{-1})^2},\quad
\mathbb{C}\cdot x_1=\varphi^1+\e\frac{q+q^{-1}}{(q-q^{-1})
^2}x_1,\quad \mathbb{C}\cdot x_2=q^{-1}\varphi_1+
\e\frac{q+q^{-1}}{(q-q^{-1}) ^2}x_2,
$$
$$
\mathbb{C}\cdot x_3=q^{-2}\varphi_1+\e\frac{q+q^{-1}}{(q-q^{-1})
^2}x_3,\quad \mathbb{C}\cdot
x_4=q^{-3}\varphi_1+\e\frac{q+q^{-1}}{(q-q^{-1}) ^2}x_4,
$$
where
$$
\varphi_1=\e(q^3x_1+q^2x_2+qx_3+x_4),
$$
$$
\mathbb{C}\cdot
x_1x_2=\e\left(q+q^3+\frac{q+q^{-1}}{(q-q^{-1})^2}\right)
x_1x_2+\e(q^2x_1x_3+qx_1x_4+qx_2x_3+x_2x_4),
$$
\be
\mathbb{C}\cdot x_1x_3=\e
q^2x_1x_2+\e\left(2q+\frac{q+q^{-1}}{(q-q^{-1})^2}\right)
x_1x_3+\e(x_1x_4+x_2x_3+x_3x_4), \ee
$$
\mathbb{C}\cdot
x_1x_4=\e(qx_1x_2+x_1x_3)+\e\left(q+q^{-1}+\frac{q+q^{-1}}
{(q-q^{-1})^2} \right)x_1x_4+\e(x_2x_4+q^{-1}x_3x_4),
$$
$$
\mathbb{C}\cdot x_2x_3=\e(qx_1x_2+x_1x_3)+\e\left(q+q^{-1}+
\frac{q+q^{-1}}{(q-q^{-1})^2} \right)x_2x_3+\e(x_2x_4+
q^{-1}x_3x_4),
$$
$$
\mathbb{C}\cdot
x_2x_4=\e(x_1x_2+x_1x_4+x_2x_3)+\e\left(2q^{-1}+\frac
{q+q^{-1}}{(q-q^{-1})^2}\right)x_2x_4+\e q^{-2}x_3x_4,
$$
$$
\mathbb{C}\cdot
x_3x_4=\e(x_1x_3+q^{-1}x_1x_4+q^{-1}x_2x_3+q^{-2}x_2x_4)+\e
\left(q^{-3}+q^{-1}+\frac{q+q^{-1}}{(q-q^{-1})^2}\right)x_3x_4,
$$
$$
\mathbb{C}\cdot x_1x_2x_3=\varphi_3,\;\; \mathbb{C}\cdot
x_1x_2x_4=q^{-1}\varphi_3,\;\; \mathbb{C}\cdot
x_1x_3x_4=q^{-2}\varphi_3,\;\; \mathbb{C}\cdot
x_2x_3x_4=q^{-3}\varphi_3,\;\; \mathbb{C}\cdot x_1x_2x_3x_4=0,
$$
where
$$
\varphi_3=\e(q^3x_1x_2x_3+q^2x_1x_2x_4+qx_1x_3x_4+x_2x_3x_4).
$$
It follows from these relations that Casimir operator has three
different eigenvalues:
$$
c_1=\e\frac{q^5+q^{-5}}{(q-q^{-1})^2},
$$
in maximally "symmetric" sector:
$$
\{\varphi_0,\ \varphi_1,\ \varphi_2,\ \varphi_3,\ \varphi_4\},
$$
where $\varphi_0=1$, $\varphi_2=q^4x_1x_2+q^3x_1x_3+q^2x_1x_4+
q^2x_2x_3+qx_2x_4+ x_3x_4$, $\varphi_4=x_1x_2x_3x_4$
$$
c_2=\e\frac{q^3+q^{-3}}{(q-q^{-1})^2},
$$
on vectors
$$
\{(x_1-qx_2),\; (x_1-q^2x_3),\; (x_1-q^3x_4),\; (x_1x_4- x_2x_3),
$$
$$
(x_1x_2+(q^{-1}-q)x_1x_3+( q^{-2}-1)x_1x_4-x_3x_4),\;
(x_1x_3+(q^{-1}-q)x_1x_4-x_2x_4),
$$
$$
(x_1x_2x_3-qx_1x_2x_4),\; (x_1x_2x_3-q^2x_1x_3x_4),\;
(x_1x_2x_3-q^3x_2x_3x_4)\}
$$
and
$$
c_3=\e\frac{q+q^{-1}}{(q-q^{-1})^2},
$$
on vectors
$$
\{(x_1x_2+(q^{-1}-q)x_1x_3-x_1x_4-x_2x_3+q^2x_3x_4),\;\;
(x_1x_3-qx_1x_4-qx_2x_3+q^2x_2x_4)\}.
$$
These eigenvalues become degenerate just for the same values of
$q$ as considered above spin 1 $\times$ spin 1 case. Indeed tensor
product of two spin $\frac 12$ spaces differs from spin 1 by
trivial one-dimensional space corresponding to spin zero. However
in this case values $q=\pm i$ are allowed too. Then one can
establish relations
$$
e^2=0=f^2,\qquad k^2=1,
$$
and
$$
c_1=c_2=c_3=0,
$$
on $P_{\frac 12 \frac 12 \frac 12 \frac 12}$. One can see that for
$q=\pm i$ eigenvectors of Casimir operator $\mathbb{C}$ do not
longer form a basis in $P_{\frac 12 \frac 12 \frac 12 \frac 12}$:
in sectors linear and trilinear with respect to $x_i$ only three
vectors from four ones are independent, while in bilinear sector
one has four independent vectors instead of six. In this way one
obtains that after appropriate completion the set of eigenvectors
of Casimir, it will take block-diagonal form with four $4\times 4$
blocks (\ref{c4}), i.e. tensor product of two indecomposable
representations decomposes into direct sum of indecomposable ones:
$I^{(4)}\otimes I^{(4)}=I^{(4)}\oplus I^{(4)} \oplus I^{(4)}\oplus
I^{(4)}$.
\section{Conclusion and summary}
Here we summarize some conclusions which follow from the
considerations of previous section. As it is known the spin
addition law or more generally representation fusion rule when $q$
is a root of unity ($q^N=1$) is not deformed if root index $N$ is
large enough, more precisely if $m<{2\mathcal{N}}$, where $m=m_1
\times m_2 \times \ldots$, $m_i$ are dimensions of representations
in tensor product. Dimension of an irrep is not exceed
$\mathcal{N}$ ($q^{\mathcal{N}}=\pm 1$). Tensor product of irreps
decomposes into direct sum of irreps and indecomposable
representations of dimension $2\mathcal{N}$ if dimension of tensor
product exceed $2\mathcal{N}$. The eigenvalues of Casimir operator
$\mathbb{C}$ play the key role in this decomposition. For
remaining Casimirs one can obtain:
$f^{\mathcal{N}}=0=e^{\mathcal{N}}$ and
$k^{\mathcal{N}}=(-\e)^{\mathcal{N}}$, $\e=\pm 1$ for
representations of $\mathcal{A}$ type.

So, when $q$ is given by a root of unity there exists the maximal
value of spin $j_{max}=\frac{\mathcal{N}-1}2$.

Another notation has crucial importance for physical applications:
when deformation parameter takes exceptional value $q=\pm i$ the
fundamental two-dimensional representation appears with property
$e^2=0=f^2$. Fusion of such representation naturally leads to the
indecomposable representation with the same property. It means
that value $q=i$ which is specific for XX Heisenberg model ensures
realization of Pauli principle peculiar to free fermions.

Let us summarize the fusion rules for the cases with lowest values
of $\mathcal{N}$, $\mathcal{N}=2,3$, considered above and give
decompositions of tensor products of all the allowed classic-like
irreps and indecomposable representations arising here. From the
discussions of previous section complete fusion rules are followed
for the tensor products of spin $\frac{1}{2}$ and $1$ irreps.

The case $\mathcal{N}=2$. The only $\mathcal{A}$ type irrep is
one-half spin $(\frac{1}{2})$ (besides of the one-dimensional zero
spin representation, on which all generators act trivially, and
quadric Casimir is 0), and from the fusion emerges one
four-dimensional indecomposable representation in accordance to
~\cite{Ar1}.

\bea \nn \frac{1}{2}\otimes \frac{1}{2}&=&I^{(4)},\\
\frac{1}{2}\otimes I^{(4)}= \bigoplus^{4}\frac{1}{2},&&
I^{(4)}\otimes I^{(4)}=\bigoplus^{4} I^{(4)}. \eea

 From these
relations follows a general rule

\bea\nn \bigotimes^{2n}\frac{1}{2}=\bigoplus^{k} I^{(4)},&\qquad
k=2^{2(n-1)},
\\\nn
\bigotimes^{2n+1}\frac{1}{2}=\bigoplus^{k} \frac{1}{2},&\qquad
k=2^{2n},
\\\nn
\bigotimes^{n} I^{(4)}= \bigoplus^{k} I^{(4)},&\qquad k=4^{(n-1)},
\\\nn
\bigotimes^{n} I^{(4)}\bigotimes^{2r}\frac{1}{2}= \bigoplus^{k}
I^{(4)},&\qquad k=4^{(n+r-1)},
\\
\bigotimes^{n} I^{(4)}\bigotimes^{2r+1}\frac{1}{2}= \bigoplus^{k}
\frac{1}{2},&\qquad k=4^{(n+r)}. \eea

For the case $\mathcal{N}=3$ the $\mathcal{A}$ type irreps are
three - with spins zero, one-half and one:
$(0)$,$(\frac{1}{2})$,$(1)$, and from their fusions two
six-dimensional indecomposable representations are arising: $
{Ind}_{\mathcal{A}}(j=0), Ind_{\mathcal{A}}(j=1)$ in the
classification of ~\cite{Ar1}. The fusion rules are

\bea\nn \frac{1}{2}\otimes \frac{1}{2}=1 \oplus 0,
\\\nn
\frac{1}{2}\otimes 1= I^{(6)}_1,
\\\nn
1 \otimes 1= I^{(6)}_2 \oplus 1,
\\\nn
\frac{1}{2}\otimes I^{(6)}_1=I^{(6)}_2 \oplus 1\oplus 1,
\\\nn
\frac{1}{2}\otimes I^{(6)}_2=I^{(6)}_1 \oplus 1\oplus 1,
\\\nn
1 \otimes I^{(6)}_1=1 \otimes I^{(6)}_2=\bigoplus^{2}I^{(6)}_1
\bigoplus^{2} 1,
\\
I^{(6)}_{1,2} \otimes I^{(6)}_{1,2}=\bigoplus^{2}I^{(6)}_1
\bigoplus^{2}I^{(6)}_2 \bigoplus^{4} 1. \eea

The generalization for the higher tensor products is obvious, all
they consist of both of spin-irreps and indecomposable
representations. For illustration let us draw for small values of
$\mathcal{N}$, $\mathcal{N}=2,3$, the extended Bratteli diagrams
(the decomposition rules for the tensor products of n copies of
similar representations) for both of irreducible and
indecomposable representations.

The tensor product of the finite dimensional representations of
$s\ell_q(2)$ is reduced into a linear combination
 \be \underbrace{V_i \otimes V_i
\otimes \cdot\cdot\cdot \otimes V_i}_{n}=\sum_{k}w^{ki}_{n}V_{k}.
\ee

Here the $w^{ki}_{n}$ are the multiplicities of the $V_{k}$
representations (irreducible and indecomposable representations).
In the Bratteli diagrams (see figures) these numbers are
consistent with the numbers of paths coming to the respective
representations (dots in the figures) from the origin. The
diagrams for the $s\ell_q(2)$ representations contain multiple
links of $r$ times (in cases $q^4=1, q^3=\pm 1$, the $r=2,4$). The
path which is passed
such link, must be multiplied by $r$. 
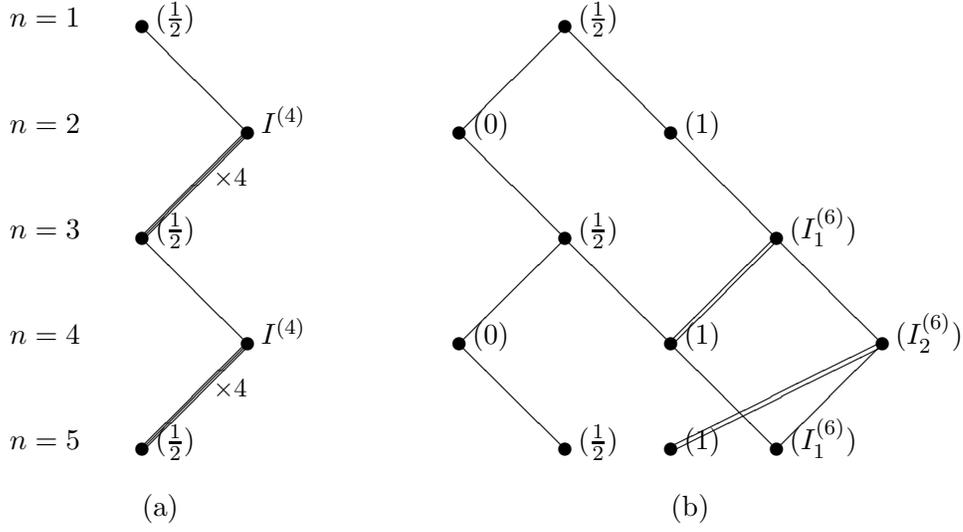
\begin{figure}[h]
\unitlength=10pt
\begin{picture}(100,20)(-1,0)
\multiput(8,20)(0,-8){2}{\line(1,-1){4}}
\multiput(8,20)(0,-8){3}{\circle*{0.5}}
\multiput(12,16)(0,-8){2}{\circle*{0.5}} {\linethickness{1.5pt}
\multiput(12,16)(0,-8){2}{\line(-1,-1){4}}
\multiput(11.9,16)(0,-8){2}{\line(-1,-1){4}}
\multiput(12.1,16)(0,-8){2}{\line(-1,-1){4}}}
\multiput(8.5,20)(0,-8){3}{$(\frac{1}{2})$}
\multiput(12.5,16)(0,-8){2}{$I^{(4)}$}
\multiput(10.7,14)(0,-8){2}{\small{$\times 4$}} \put(3,20){$n=1$}
\put(3,16){$n=2$} \put(3,12){$n=3$} \put(3,8){$n=4$}
\put(3,4){$n=5$}

\multiput(24,20)(4,-4){3}{\line(1,-1){4}}
\multiput(24,20)(0,-8){2}{\line(-1,-1){4}}
\multiput(20,16)(0,-8){2}{\line(1,-1){4}}{\linethickness{1pt}
\put(35.8,8){\line(-2,-1){8}} \put(36.2,8){\line(-2,-1){8}}
\put(31.9,12){\line(-1,-1){4}}\put(32.1,12){\line(-1,-1){4}}}
\multiput(36,8)(-4,-4){1}{\line(-1,-1){4}}
\multiput(24,12)(4,-4){2}{\line(1,-1){4}}
\multiput(24.5,20)(0,-8){3}{$(\frac{1}{2})$}
\multiput(20.5,16)(0,-8){2}{$(0)$}
\multiput(28.5,16)(0,-8){2}{$(1)$}
\multiput(32.5,12)(0,-8){2}{$(I^{(6)}_1)$}
\put(36.5,8){$(I^{(6)}_2)$} \put(28.5,4){$(1)$}
\multiput(24,20)(0,-8){3}{\circle*{0.5}}
\multiput(20,16)(0,-8){2}{\circle*{0.5}}
\multiput(28,16)(0,-8){2}{\circle*{0.5}}
\multiput(32,12)(0,-8){2}{\circle*{0.5}} \put(36,8){\circle*{0.5}}
\put(28,4){\circle*{0.5}}
\put(8,1.5){(a)}\put(28,1.5){(b)}
\end{picture}
\label{BR1}\vspace{-1cm} \caption{Bratteli diagrams for the irreps
$\frac{1}{2}$ in cases: a) $\mathcal{N}=2,\quad b)
\mathcal{N}=3$.}
\end{figure}
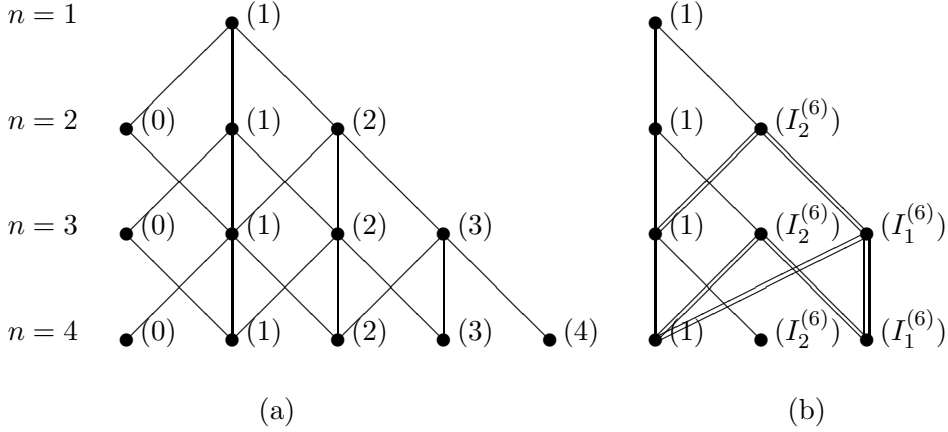
\begin{figure}[h]
\unitlength=10pt
\begin{picture}(100,20)(-1,0)
\put(2.5,20){$n=1$} \put(2.5,16){$n=2$} \put(2.5,12){$n=3$}
\put(2.5,8){$n=4$} \multiput(11,20)(0,-4){3}{\line(1,-1){4}}
\multiput(11,20)(0,-4){3}{\line(-1,-1){4}}
\put(11,20){\line(0,-1){12}} \put(15,16){\line(0,-1){8}}
\multiput(7,16)(0,-4){2}{\line(1,-1){4}}
\multiput(15,16)(0,-4){2}{\line(-1,-1){4}}
\multiput(15,16)(0,-4){2}{\line(1,-1){4}}
\put(19,12){\line(-1,-1){4}}\put(19,12){\line(0,-1){4}}\put(19,12){\line(1,-1){4}}
\multiput(7.5,16)(0,-4){3}{$(0)$}
\multiput(11.5,20)(0,-4){4}{$(1)$}
\multiput(15.5,16)(0,-4){3}{$(2)$}
\multiput(19.5,12)(0,-4){2}{$(3)$} \put(23.5,8){$(4)$}
\multiput(7,16)(0,-4){3}{\circle*{0.5}}
\multiput(11,20)(0,-4){4}{\circle*{0.5}}
\multiput(15,16)(0,-4){3}{\circle*{0.5}}
\multiput(19,12)(0,-4){2}{\circle*{0.5}} \put(23,8){\circle*{0.5}}
\put(12,5){(a)}
\multiput(27,20)(0,-4){3}{\line(0,-1){4}}
\multiput(27,20)(0,-4){3}{\line(1,-1){4}}
\multiput(30.9,16)(0,-4){2}{\multiput(0,0)(0.2,0){2}{\line(1,-1){4}}}
\multiput(30.9,16)(0,-4){2}{\multiput(0,0)(0.2,0){2}{\line(-1,-1){4}}}
\multiput(34.9,12)(0,-4){1}{\multiput(0,0)(0.2,0){2}{\line(0,-1){4}}}
\multiput(34.8,12)(0,-4){1}{\multiput(0,0)(0.4,0){2}{\line(-2,-1){8}}}
\multiput(27.5,20)(0,-4){4}{$(1)$}
\multiput(31.5,16)(0,-4){3}{$(I^{(6)}_2)$}
\multiput(35.5,12)(0,-4){2}{$(I^{(6)}_1)$}
\multiput(27,20)(0,-4){4}{\circle*{0.5}}
\multiput(31,16)(0,-4){3}{\circle*{0.5}}
\multiput(35,12)(0,-4){2}{\circle*{0.5}} \put(32,5){(b)}
\end{picture}
\vspace{-2cm} \caption{Bratteli diagrams for the irreps $1$ in
cases: a) q is not root of unity,\quad b) $\mathcal{N}=3$.}
\label{BR2}
\end{figure}

The next steps of the towers ($n\geq 6$ in Fig.1, and $n\geq 5$ in
Fig.2(b) and Fig.3), contain the same representations already
appeared for the lower n-s. The multiple links are drawn either by
$r$ parallel lines or by thick lines with label $(\times r)$. For
comparison in Fig.2a we represent the case for $s\ell(2)$ algebra.
As it is expected \cite{Ar1,PS} the fusions of the $\mathcal{A}$
type representations form closed ring.

We can do some remark about values of $\mathcal{N}$ higher than 3.
The maximal allowed spin representation with dimension
$\mathcal{N}$ has spin $j_{max}=\frac{\mathcal{N}-1}{2}$. The
tensor product of $j_{max}$ with $\frac{1}{2}$ is an
indecomposable representation with dimension $2\mathcal{N}$:

\be j_{max}\otimes \frac{1}{2}=I^{2\mathcal{N}}_1, \ee
\be
I^{2\mathcal{N}}_1\otimes \frac{1}{2}=j_{max}\otimes
\frac{1}{2}\otimes \frac{1}{2}=j_{max}\otimes (1 \oplus
0)=j_{max}\oplus (j_{max}\otimes 1). \ee
 We expect that for
general case also $(j_{max}\otimes 1)$ expands to the sum of
$j_{max}$ and another indecomposable representation
$I^{2\mathcal{N}}_2$. \bea j_{max}\otimes 1=j_{max} \oplus
I^{2\mathcal{N}}_2.
 \eea

So the representation with maximum spin together with
indecomposable representations appears in decomposition of the
tensor product of an indecomposable representation with any
other~\cite{dck} . By definition \cite{PS} these are states with
zero q-dimension (for irreps $\dim_q\rho_j=[2j+1]_q$).

\begin{figure}[t] \unitlength=10pt
\begin{picture}(100,20)(-1,0)
\put(2.5,20){$n=1$} \put(2.5,16){$n=2$} \put(2.5,12){$n=3$}
\put(2.5,8){$n=4$} \multiput(10.85,20)(0.1,0){4}{\line(0,-1){12}}
\multiput(9,20)(0,-4){4}{\scriptsize $I^{(4)}$}
\multiput(11.5,18)(0,-4){3}{\scriptsize $\times 4$}
\multiput(11,20)(0,-4){4}{\circle*{0.5}} \put(12,5){(a)}
\multiput(26.9,16)(8,0){2}{\multiput(0,0)(0.2,0){2}{\line(0,-1){8}}}
\multiput(30.9,20)(0.2,0){2}{\line(0,-1){12}}
\multiput(30.9,20)(0,-4){3}{\multiput(0,0)(0.2,0){2}{\line(1,-1){4}}}
\multiput(30.9,20)(0,-4){3}{\multiput(0,0)(0.2,0){2}{\line(-1,-1){4}}}
\multiput(31,20)(0,-4){3}{\line(-1,-1){4}}
\multiput(34.9,16)(0,-4){2}{\multiput(0,0)(0.2,0){2}{\line(-1,-1){4}}}
\multiput(34.9,16)(0,-4){2}{\multiput(0,0)(0.2,0){2}{\line(-2,-1){4}}}
\multiput(34.9,16)(0,-4){2}{\multiput(0,0)(0.2,0){2}{\line(-2,-1){8}}}
\multiput(35,16)(0,-4){2}{\line(-2,-1){8}}
\multiput(34.8,16)(0,-4){2}{\multiput(0,0)(0.4,0){2}{\line(-2,-1){8}}}
\multiput(26.8,16)(0,-4){2}{\multiput(0,0)(0.4,0){2}{\line(2,-1){8}}}
\multiput(25.5,16)(0,-4){3}{\scriptsize$(1)$}
\multiput(31.5,20)(0,-4){4}{\scriptsize $(I^{(6)}_{2})$}
\multiput(35.5,16)(0,-4){3}{\scriptsize$(I^{(6)}_{1})$}
\multiput(28,18.2)(0,-4){3}{\scriptsize $\times 4$}
\multiput(29,12.5)(0,-4){2}{\scriptsize $\times 4$}
\multiput(27,16)(0,-4){3}{\circle*{0.5}}
\multiput(31,20)(0,-4){4}{\circle*{0.5}}
\multiput(35,16)(0,-4){3}{\circle*{0.5}} \put(32,5){(b)}
\end{picture}
\vspace{-2cm} \caption{Extended Bratteli diagrams for fusions of
indecomposable representations $(I)^n$ in cases: a)
$\mathcal{N}=2$,\quad b) $\mathcal{N}=3$.} \label{BR2}
\end{figure}
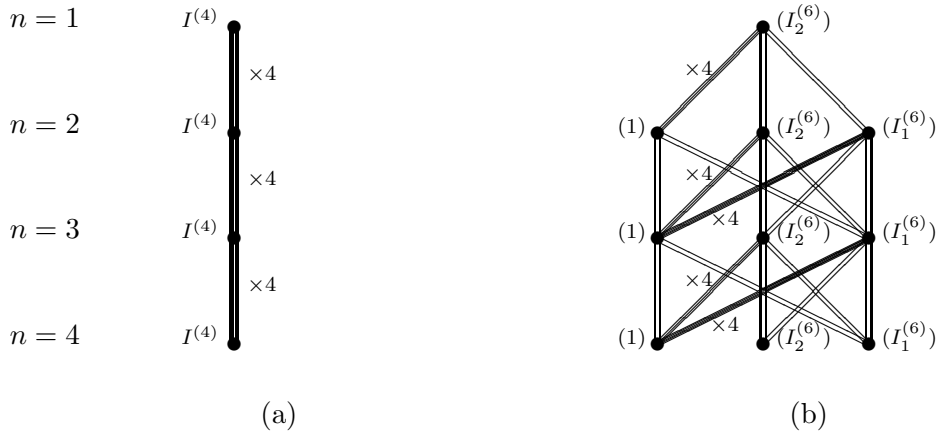

\section{\large Acknowledgments}
The work was partially supported by the Volkswagen Foundation of
Germany and INTAS grant No 03-51-5460.

Authors would like to thank A. Sedrakyan and D. Arnaudon for
helpful notations.

\end{document}